\documentclass{aa}  

\usepackage{graphicx}
\usepackage{txfonts}
\newcommand{\SFR}{\textrm{M}$_{\odot}$~\textrm{yr}$^{-1}$}
\newcommand{\kms}{$\textrm{km~s$^{-1}$}$}
\newcommand{\Ha}{H$\alpha$} 

\begin{document}

   \title{Star formation in outer rings of S0 galaxies.}

   \subtitle{I. NGC 6534 and MCG 11-22-015.}

   \author{O. Sil'chenko
          \inst{1}
          \and
          I. Kostiuk
          \inst{2}
          \and
          A. Burenkov
          \inst{2}
          \and
          H. Parul\inst{3}
          }

   \institute{Sternberg Astronomical Institute of the Lomonosov Moscow
             State University, University av. 13, 119991 Russia\\
             \email{olga@sai.msu.su}
          \and
             Special Astrophysical Observatory
             of the Russian Academy of Sciences,
             Nizhnij Arkhyz, 369167 Russia\\
              \email{kostiuk@sao.ru,ban@sao.ru}
         \and
              Department of Physics and Astronomy, University of Alabama,
              P.O. Box 870324, Tuscaloosa, AL 35487-0324, USA\\
              \email{hparul@crimson.ua.edu}
             }

   \date{Received  .., 2018; accepted .., 2018}

% \abstract{}{}{}{}{} 
% 5 {} token are mandatory
 
  \abstract
   {}
   {Though S0 galaxies are usually thought to be `red and dead', they often
   demonstrate star formation organized in ring structures. We try
   to clarify the nature of this phenomenon and its difference from star
   formation in spiral galaxies. Two early-type galaxies with outer rings, 
   NGC~6534 and MCG~11-22-015, are selected to be studied.}
   {After inspecting the gas excitation in the rings using the Baldwin-Phillips-Terlevich method \citep{bpt},
   we estimate the star formation rates (SFR) in the two outer rings of our galaxies by
   using several SFR indicators derived from narrow-band photometry in the H$\alpha$ emission
   line and archival GALEX ultraviolet images of the galaxies.}
   {The ionized gas is excited by young stars in the ring of NGC 6534 and partly
   by shocks -- in MCG 11-22-015. The oxygen abundances in the HII regions of
   the rings are close to solar. The derived SFRs allow to qualitatively restore star
   formation histories (SFH) in the rings: in NGC~6534 the SFH is flat during
   the last 100-200~Myr, and in MCG~11-22-015 the star formation has started only
   a few Myr ago. We suggest that the rings in NGC~6534 and MCG~11-22-015 have different
   natures: the former is a resonant one supplied with gas perhaps through tidal effects,
   and the latter has been produced by a satellite accretion. Recent outer gas accretion is implied in both cases.}
   {}


   \maketitle
%
%-------------------------------------------------------------------

\section{Introduction}

   Early-type reddish galaxies possessing large-scale detached outer stellar rings were first 
   noted by \citet{devauc59}.
   Boris A. Vorontsov-Velyaminov, during his compilation of the `Morphological
   Catalogue of Galaxies' (MCG), had found the galaxies with large-scale rings to be so numerous
   that he proposed to introduce a special branch into the Hubble's fork to describe
   them \citep{vv60}. Recent surveys reveal outer stellar rings to be found mostly
   in S0 galaxies -- according to \citet{arrakis}, up to 50\%\ S0--S0/a galaxies 
   have outer rings. A similar fraction of ring structures in early-type disk galaxies,
   $53\pm 5$\%, was noted by \citet{nirs0s}.

   Rings are usually thought to be mostly of resonance nature: they represent
   the result of gas accumulation at Lindblad resonances of large-scale bars
   and of consequent star formation. However, not all galaxies with outer
   rings have large-scale bars; and not all S0s with outer rings
   demonstrate any gas presence and star formation in the rings. Our estimates
   based on the listing of the ARRAKIS catalogue \citep{arrakis} and on the GALEX imaging data
   reveal that about half of all outer rings in S0s have had star formation
   during the last 100-200~Myr \citep{kostuk15}. The problem of gas sources
   to feed starforming rings in S0 galaxies is still unsolved; and nothing
   is known about the star formation histories of these spectacular
   structures. In this Letter we will consider NGC~6534 and MCG~11-22-015 -- two
   early-type disk galaxies belonging to the red sequence with outer rings;
   their global properties are given in Table~\ref{global_data}.
   The optical-band rings and first spectral results for these galaxies were reported 
   by one of us long ago \citep{kostuk1975,kostuk1981}.
   Later, after the GALEX data releases, we have detected there UV-rings \citep{we_uvrings}
   with the radii corresponding to the optical rings.
   Now we will present new spectral data as well as SFR estimates obtained
   from the UV images and from our own \Ha\ narrow-band imaging.

\begin{table}
\caption[ ] {Global parameters of the galaxies}
\label{global_data}
% %\begin{center}
\begin{flushleft}
\begin{tabular}{lcc}
\hline\noalign{\smallskip}
% % & & & \\
Galaxy & NGC~6534 & MCG~11-22-015  \\
\hline
$R_{25}$, arcsec (LEDA) &  28   &  24  \\
$R_{ring}$ (outer edge), arcsec & $19 \times 13$ & $20\times 18$ \\
$M_H$(NED)  & --23.5 & --23.7  \\
$M_{NUV}$, (our estimate) & --17.2 & --16.1 \\
$(u-r)_0$ (SDSS/DR9) & 2.55 & 2.63  \\
$V_r $ (NED), $\mbox{km} \cdot \mbox{s}^{-1}$ &  8118 &
      8064   \\
Distance, Mpc (NED) & 110 & 109 \\
Scale, kpc per arcsec (NED) & 0.518 & 0.514 \\
{\it PA}$_{phot}$ (LEDA) & $16.5^{\circ}$ & $75^{\circ}$  \\
{\it PA}$_{ring}$ (our estimate) & $21.7^{\circ}$ & $72.5^{\circ}$ \\
\hline
\end{tabular}
\end{flushleft}
\end{table}

%--------------------------------------------------------------------
\section{Observations and star formation rate calibrations}

We exposed both galaxies with the SCORPIO reducer of the Russian 6m telescope \citep{scorpio}
operating in the imaging mode, through the narrow-band filter centered onto
the redshifted \Ha\ line (both galaxies have close radial velocities)
and in the neighbouring continuum. A spectrophotometric standard, AGK$+$81d266, was also exposed
the same night through the FN674 filter. We also used the SCORPIO reducer 
to obtain spectra. We obtained green- and red-band
medium-resolution spectra for NGC~6534 and only a red spectrum for MCG~11-22-015.
The details of the observations are given in Table~\ref{table:obs}.

To estimate the SFR over various temporal scales, we have used the GALEX images
in the NUV and FUV bands and our calibrated \Ha\ images; the formulae for
the SFR calculations have been taken from \citet{hao11sfr} and from the
review by \citet{ken_evans}.

\begin{table*}
\caption{Log of the observations with the Russian 6m telescope BTA}             % title of Table
\label{table:obs}      % is used to refer this table in the text
\centering                          % used for centering table
\begin{tabular}{c l c c c c c}        % centered columns (7 columns)
\hline\hline                 % inserts double horizontal lines
Date & Galaxy & Config/Grism or Filter & Exposure & PA(slit) & Spectral range & Seeing \\    % table heading 
\hline                        % inserts single horizontal line
Oct 07, 2016 & MCG~11-22-015 & slit/VPHG1800R & 60 min & 50\degr & 6100--7100~\AA & $1\farcs 5$ \\
Oct 07, 2016 & NGC~6534 & slit/VPHG1800R & 60 min & 20\degr & 6100--7100~\AA & $1\farcs 5$ \\
June 24, 2017 & NGC~6534 & slit/VPHG2300G & 60 min & 16.5\degr & 4800--5500~\AA & $3\farcs 3$ \\
May 28, 2017 & NGC~6534 & directimage/FN674 & 40 min & 124\degr & 6703--6763~\AA & $1\farcs 5$ \\
May 29, 2017 & NGC~6534 & directimage/SED608 & 17.5 min & 124\degr & 5976--6144~\AA & $1\farcs 8$ \\
May 28, 2017 & MCG~11-22-015 & directimage/FN674 & 30 min & 124\degr & 6703--6763~\AA & $1\farcs 5$ \\
May 29, 2017 & MCG~11-22-015 & directimage/SED608 & 15.5 min & 124\degr & 5976--6144~\AA & $1\farcs 8$ \\
May 28, 2017 & AGK$+$81d266 & directimage/FN674 & 1 min & -- & 6703--6763~\AA & $1\farcs 5$ \\
\hline                                   %inserts single line
\end{tabular}
\end{table*}

\section{The structure of NGC~6534 and MCG~11-22-015.}

   \begin{figure*}
   \centering
   \vspace{0.5cm}
   \centerline{
   \includegraphics[width=0.4\textwidth]{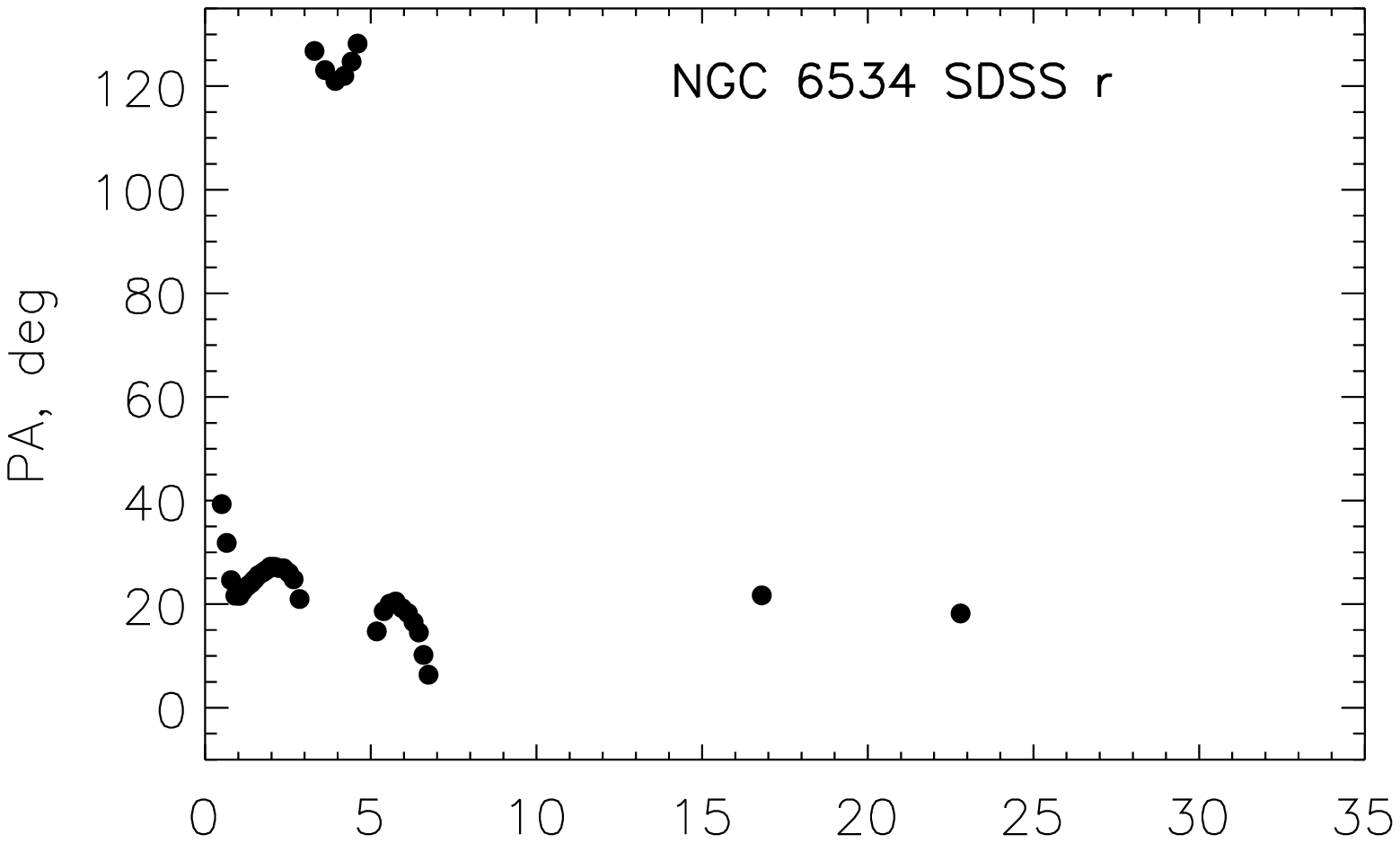}
   \includegraphics[width=0.4\textwidth]{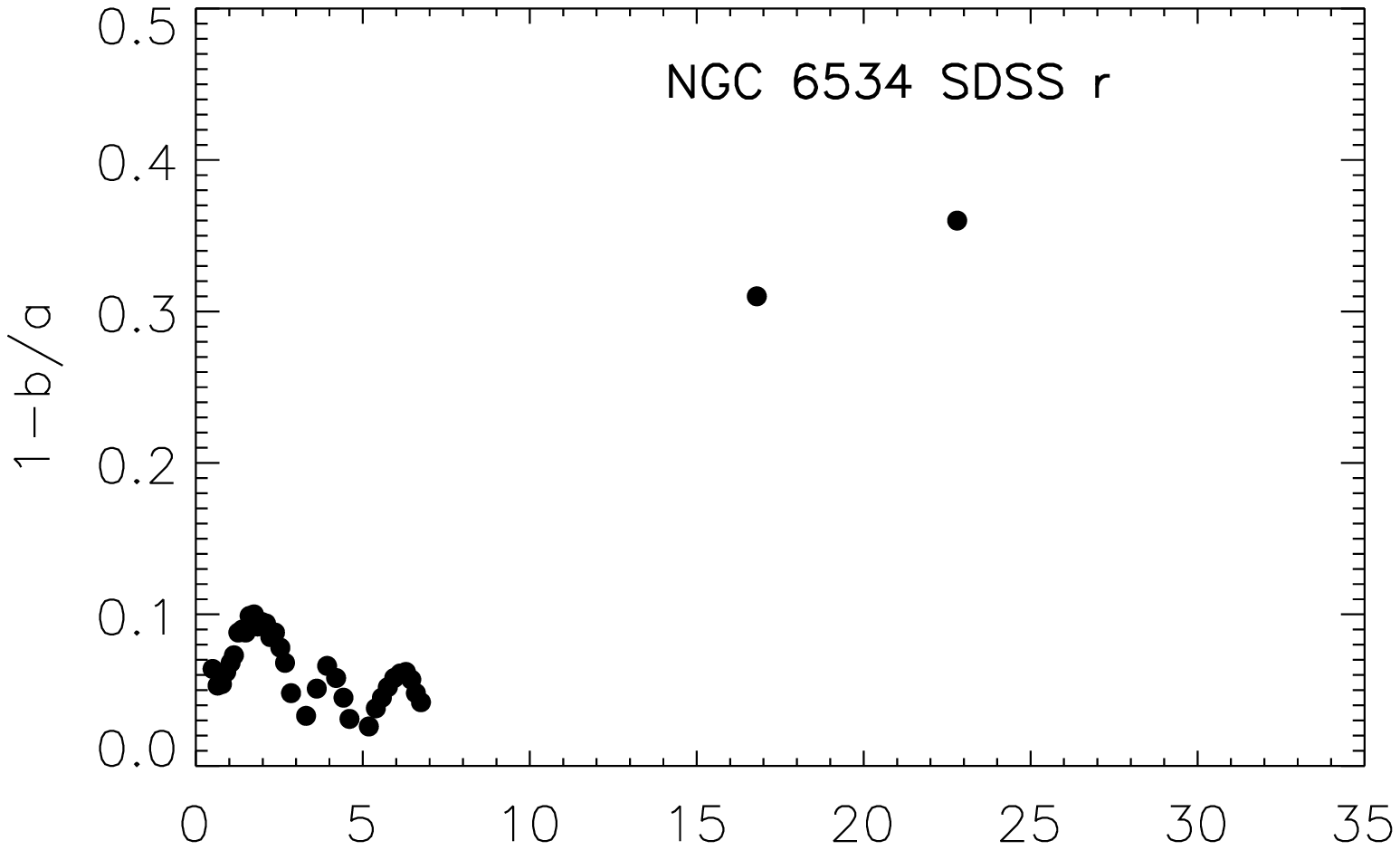}
   }
   \centerline{
   \includegraphics[width=0.45\textwidth]{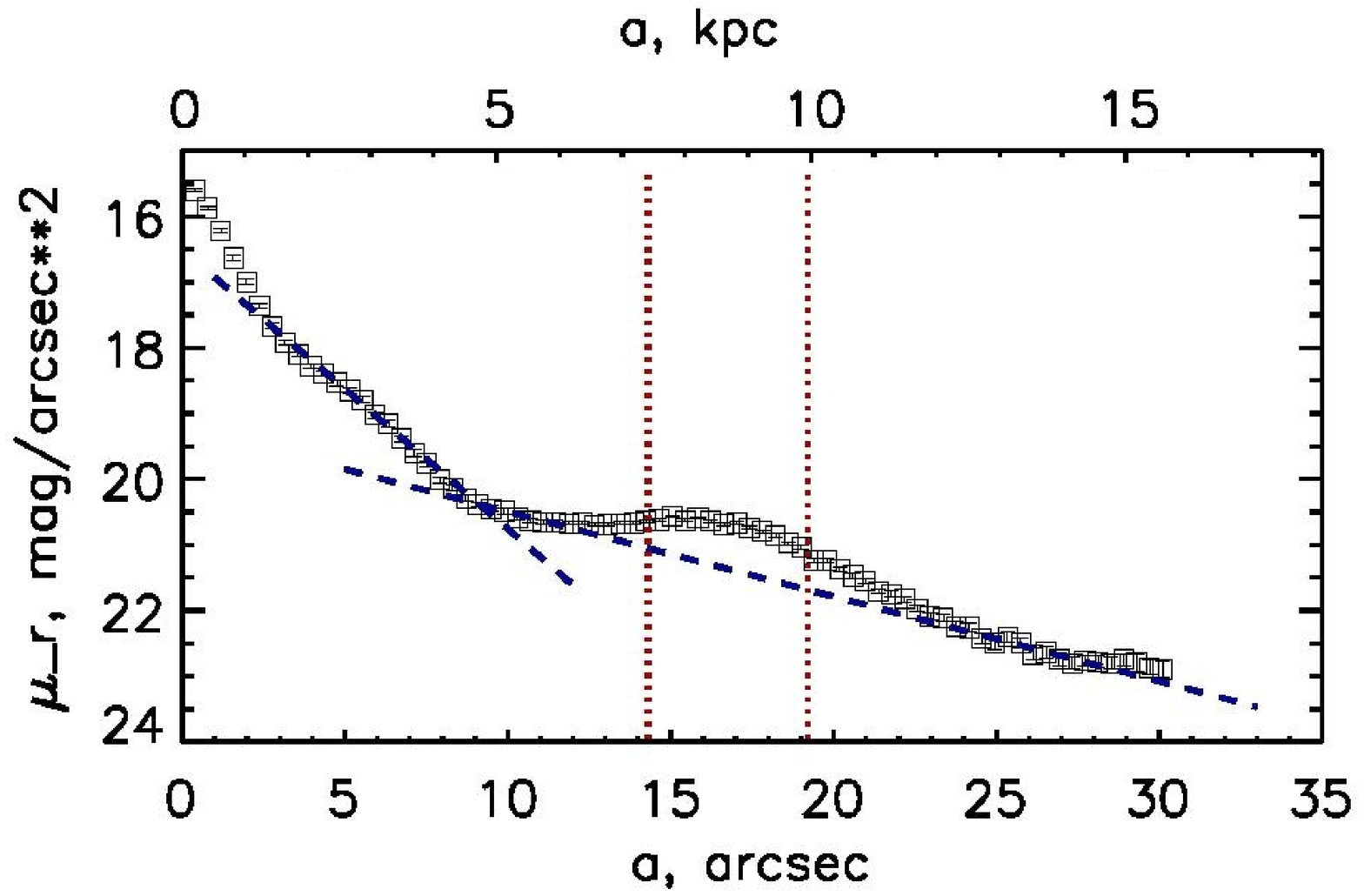}
   \includegraphics[width=0.35\textwidth]{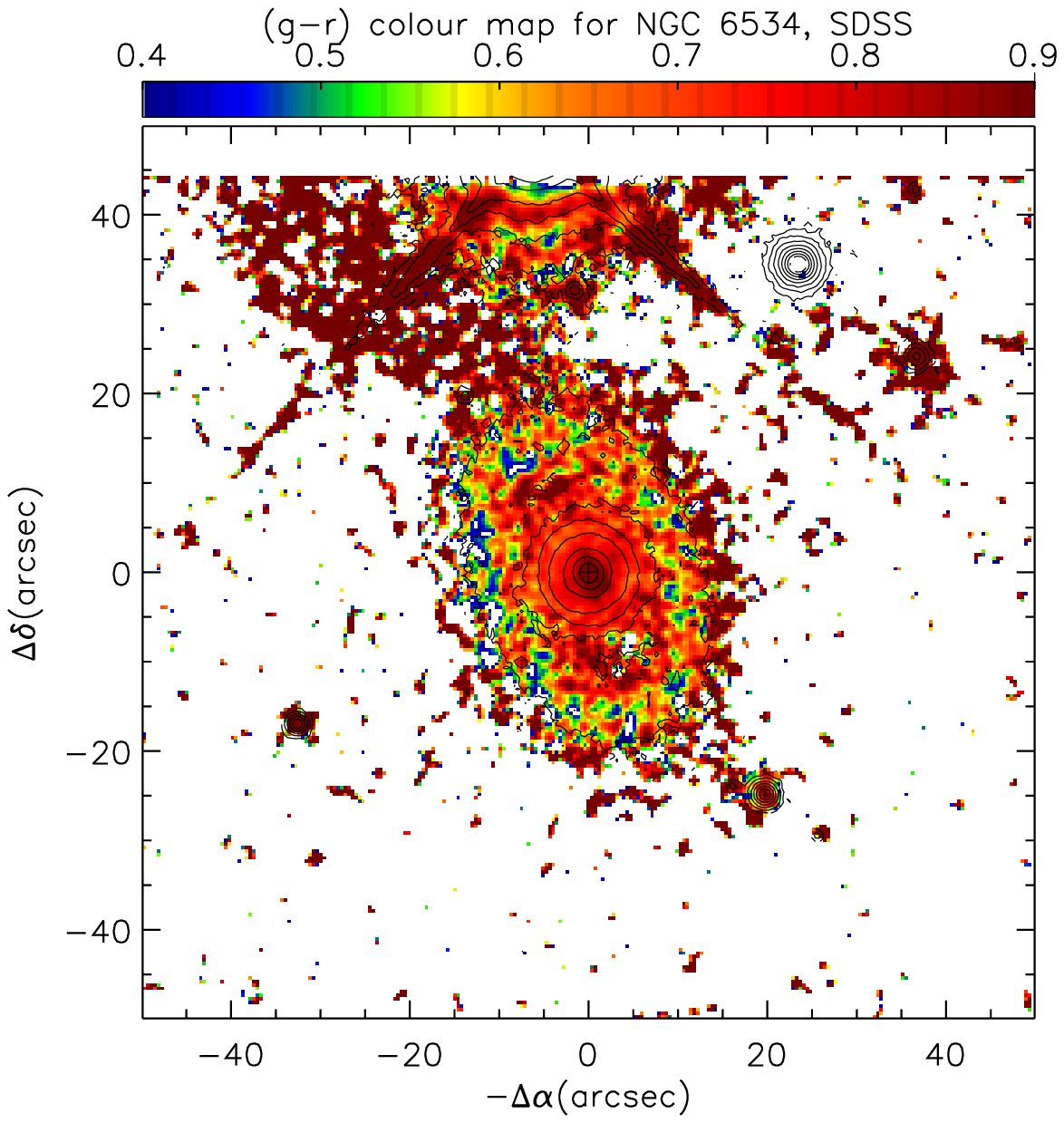} }
   \caption{SDSS-photometry of NGC~6534: the results of the
   isophote analysis ({\it upper row}), the azimuthally averaged
   surface brightness profiles ({\it bottom left}), and the colour map ({\it bottom right}).
   The fitted exponential laws are also shown. The red dotted verticals mark the borders of
   the \Ha\ ring.}
              \label{n6534ph}
    \end{figure*}

    \begin{figure*}
   \centering
   \vspace{0.5cm}
   \centerline{
   \includegraphics[width=0.4\textwidth]{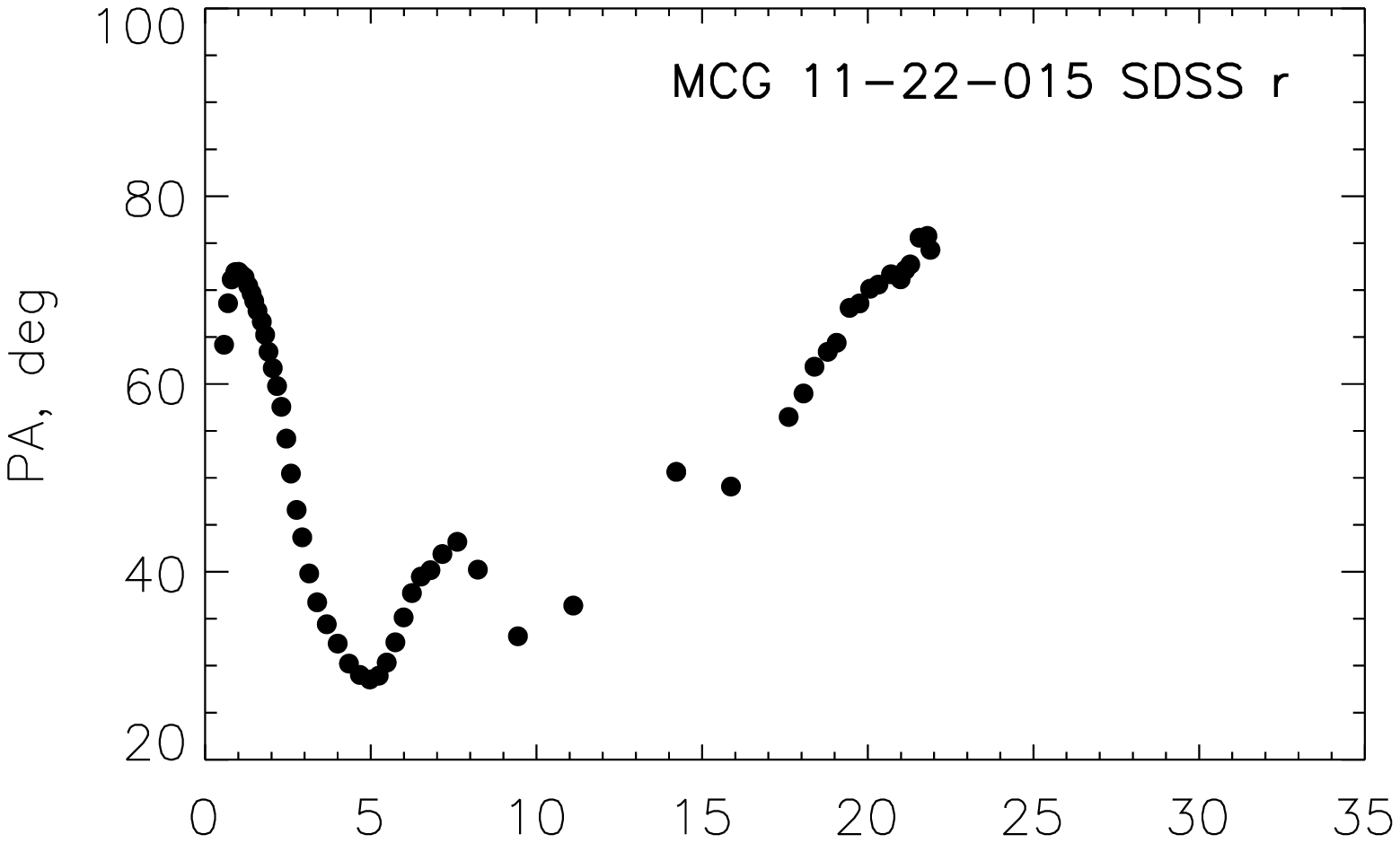}
   \includegraphics[width=0.4\textwidth]{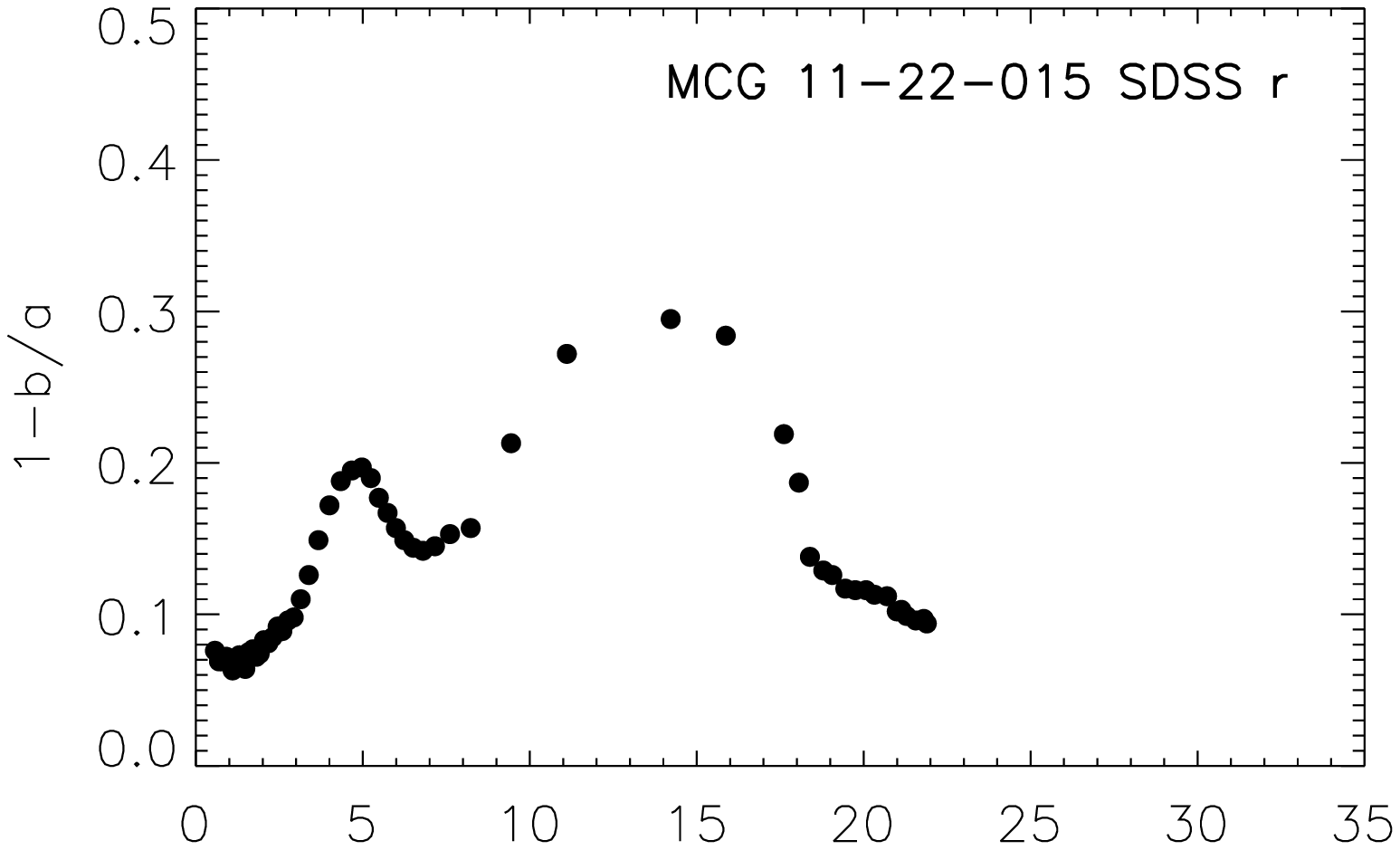}
   }
   \centerline{
   \includegraphics[width=0.45\textwidth]{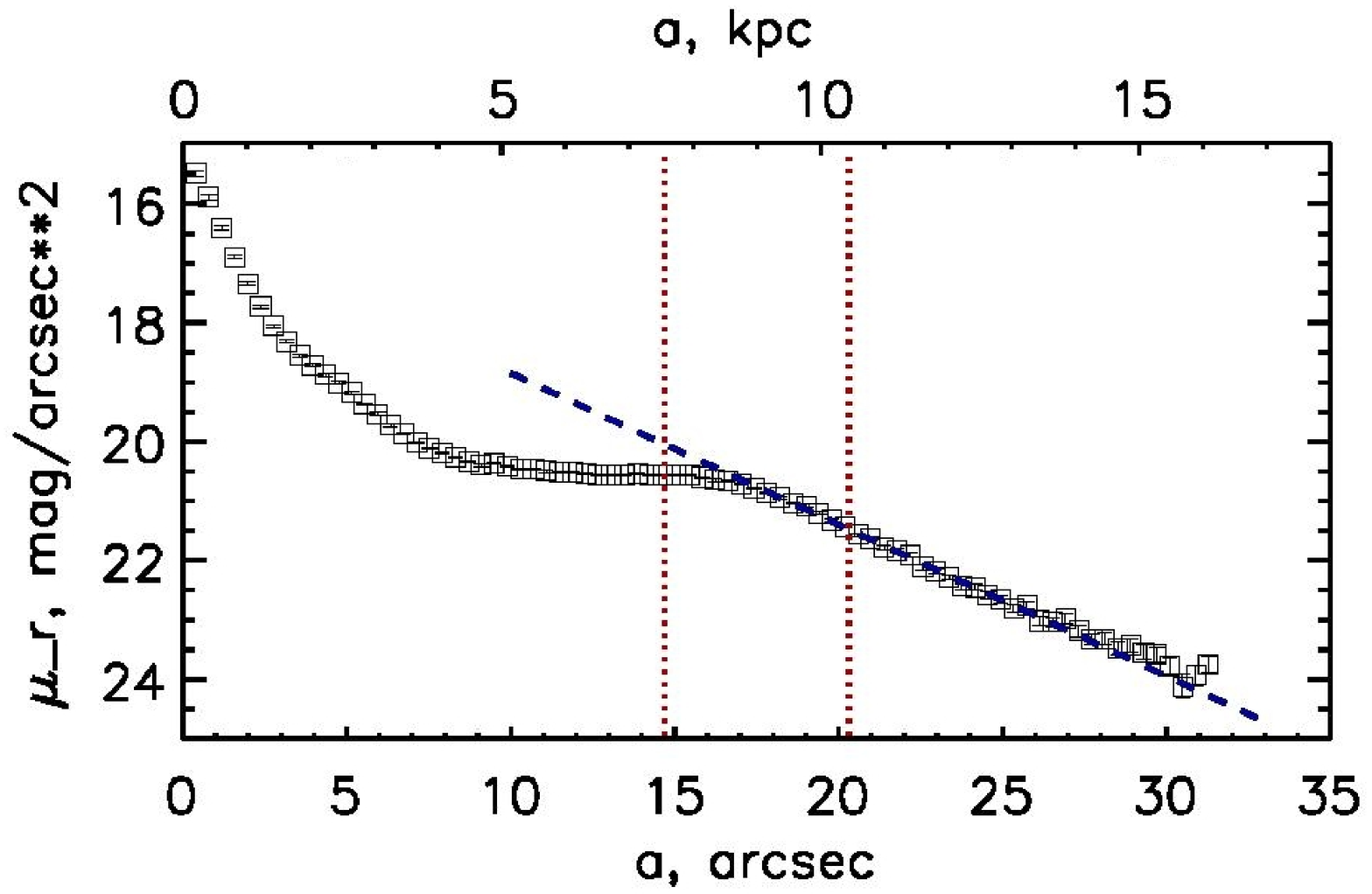}
   \includegraphics[width=0.35\textwidth]{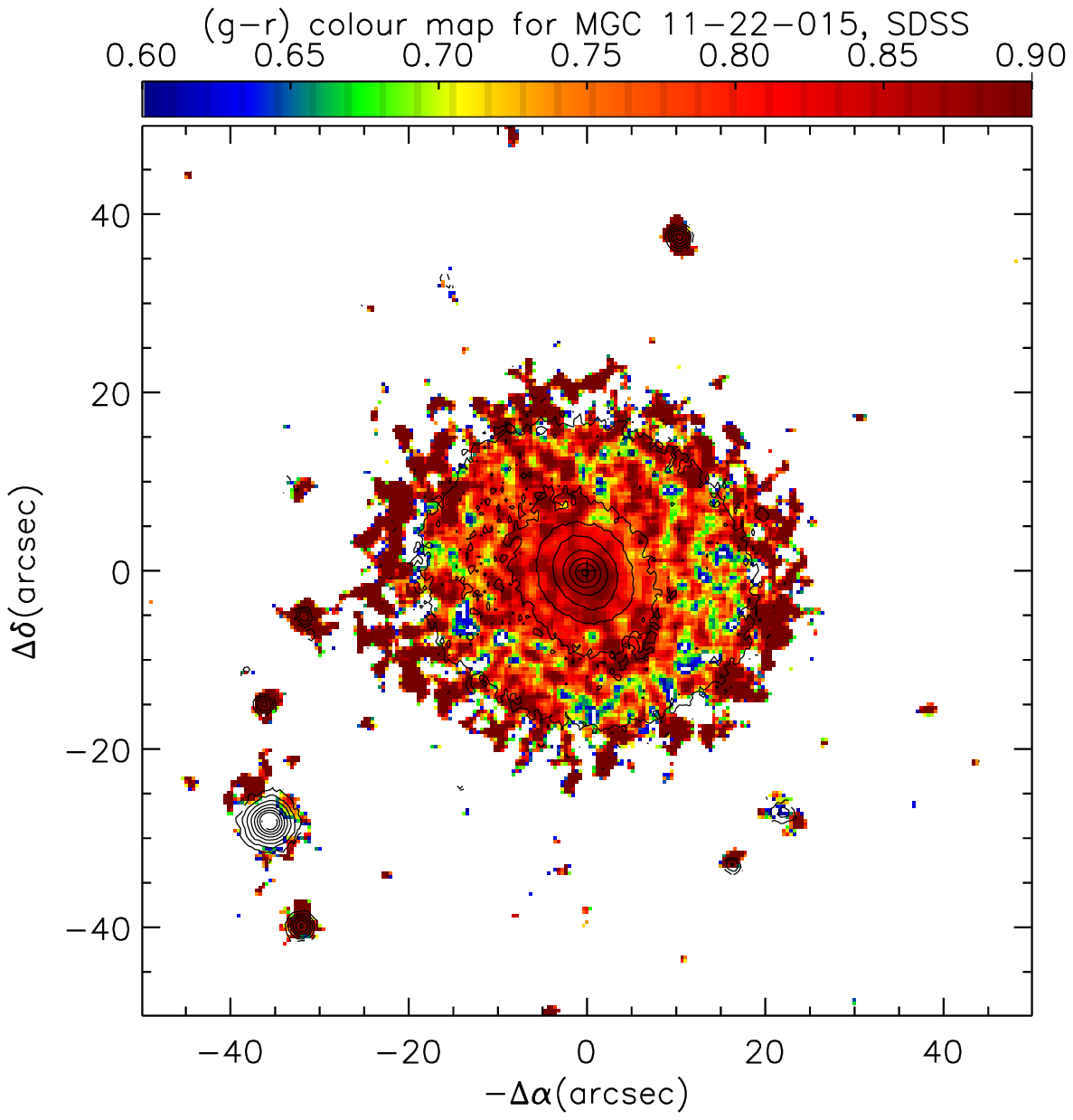} }
   \caption{SDSS-photometry of MCG~11-22-015: the results of the
   isophote analysis ({\it upper row}), the azimuthally averaged
   surface brightness profiles ({\it bottom left}), and the colour map ({\it bottom right}).
   The fitted exponential law valid over two-scalelength extension is also shown. The red dotted
   verticals mark the borders of the \Ha\ ring.}
              \label{mcgph}
    \end{figure*}

To study the large-scale structures of the galaxies, we have used the imaging
data from the SDSS/DR9. We have made an isophotal analysis, and then we have
calculated azimuthally averaged surface brightness profiles
by integrating the fluxes within elliptical annuli with running major axes and with a fixed
ellipticity and major-axis orientation determined by the outermost isophote measurements.
The results are shown in Fig.~\ref{n6534ph} and Fig.~\ref{mcgph}, together with
the colour maps $g-r$. One can see some differences in the whole structure and
ring visibility in NGC~6534 and MCG~11-22-015. NGC~6534 is moderately inclined
to our line of sight as demonstrated by the elongated
outer isophotes; the outer ring matches the ellipticity of the outermost
isophotes being evidently a round structure inside the exponential stellar disk.
However, the inner isophotes, $R\le 8\arcsec$, are very round; taking in mind
quasi-exponential surface-brightness profile of the inner part of NGC~6534 we
can suspect a pseudobulge containing an inner stellar ring slightly elongated
orthogonally with respect to the outer isophotes and to the outer
ring. MCG~11-22-015, on the contrary, has rather round outer isophotes and two
local ellipticity maxima at $R=5\arcsec$ and at $R=14\arcsec$. The latter radius
corresponds roughly to the outer ring, so if we assume that the galaxy is
seen face-on we must recognize the elliptical shape of its outer ring.
The surface-brightness profiles show the prominent ring in NGC~6534 overposed
onto the regular exponential disk ($\mu _{r,0} =19.4$, $h=9.8\arcsec$, or 5.1~kpc)
and a Type-II disk, according to the \citet{freeman} classification, in MCG~11-22-015,
with the scalelength of $4\farcs 2$, or 2.2~kpc. The elliptical ring in
MCG~11-22-015 represents a boundary between the elongated lens
and the outer exponential disk. Despite the structural differences,
the rings in both  galaxies are bluer than the underlying disks, in accordance with the
fact that the rings were initially noted as UV-bright features \citep{we_uvrings}.

\section{Corotating gas in NGC~6534}

\begin{figure}
\centering
\includegraphics[width=6cm]{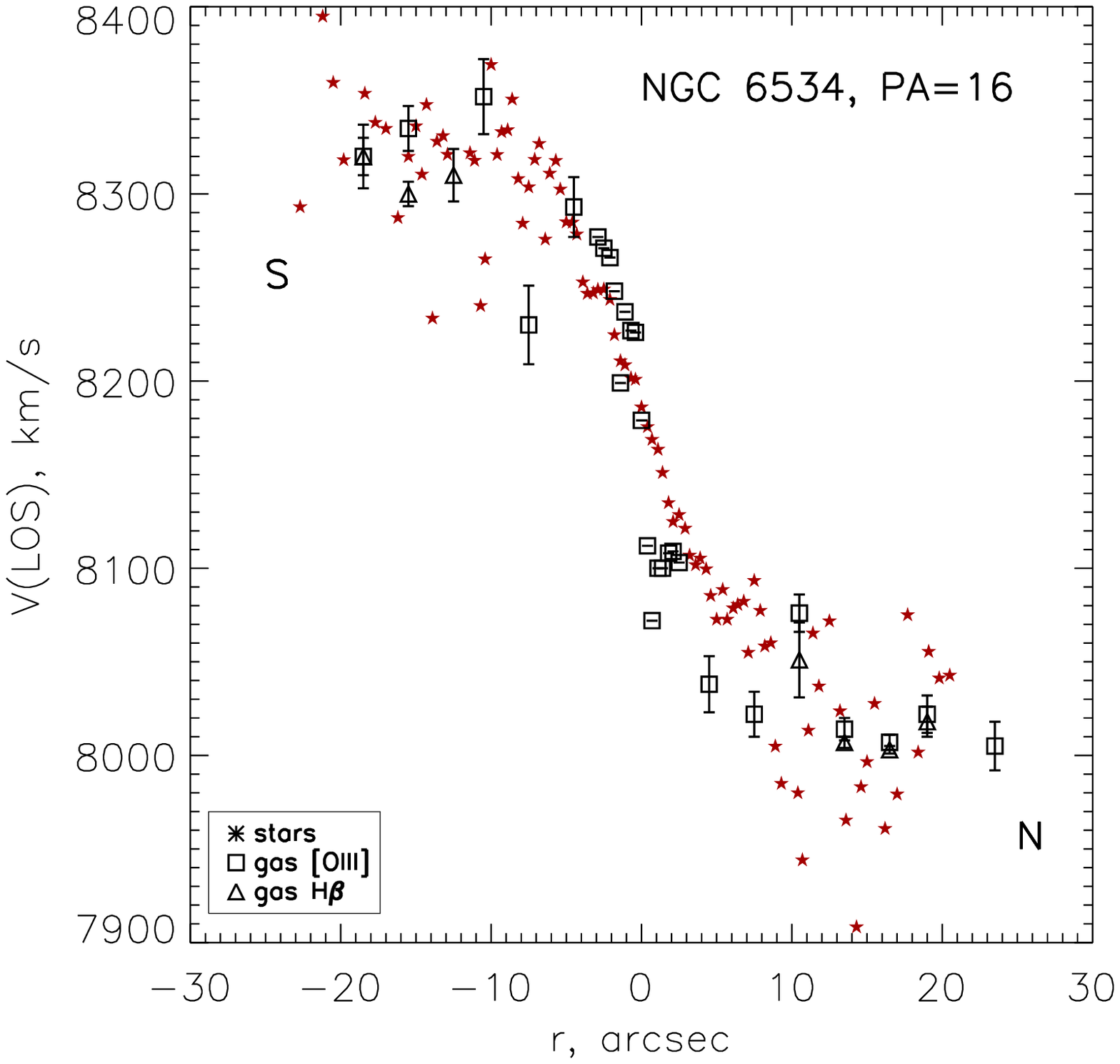}
\caption{The line-of-sight velocity profiles for the stars and ionized gas
in NGC~6534 taken along its major axis.}
\label{velprof}
\end{figure}

It is only for NGC~6534 that we have obtained the spectrum in the green with
a spectral resolution of 2~\AA, to determine the stellar and ionized-gas
rotation curves and velocity dispersion. The velocities are shown in Fig.~\ref{velprof}.
The gas in NGC~6534 rotates strictly with the stellar component, and we can
conclude that probably it is confined to the stellar-disk plane. The profile of the stellar velocity
dispersion is typical for lenticular galaxies: it has a moderate maximum in
the center, $\sigma _{*,0} =140$~\kms, and fast drop at $R>10\arcsec$ corresponding
to the onset of disk domination in the total surface brightness.

When we exposed the green spectrum of NGC~6534 with the slit at $PA=16.5\degr$,
the southern nearby galaxy, twice less luminous than NGC~6534 and also with
S0 morphology, was serendipitously caught by the slit. Its radial velocity is not
given in the NED database, so NGC~6534 is formally classified as an isolated
galaxy. From the absorption-line spectrum of the satellite we have
derived the systemic velocity of 2MASX J17560734$+$6415509,
$V_r=8124 \pm 20$~\kms. Since NGC~6534 has $V_r=8118$~\kms\ (NED),
2MASX J17560734$+$6415509 is a real and very close satellite
of NGC~6534. It is slightly bluer than NGC~6534, according to the SDSS/DR9 data;
but any signatures of ionized-gas presence are absent from the spectrum.
The projected distance between the galaxies is 75\arcsec, or only 39~kpc.
If 2MASX J17560734$+$6415509 (or LEDA 2666218) has anytime been a gas-rich
satellite, it might be a gas donor for NGC~6534.

\section{Gas excitation and oxygen abundance in the rings}

\begin{figure*}
   \centering
   \centerline{
   \includegraphics[width=0.45\textwidth]{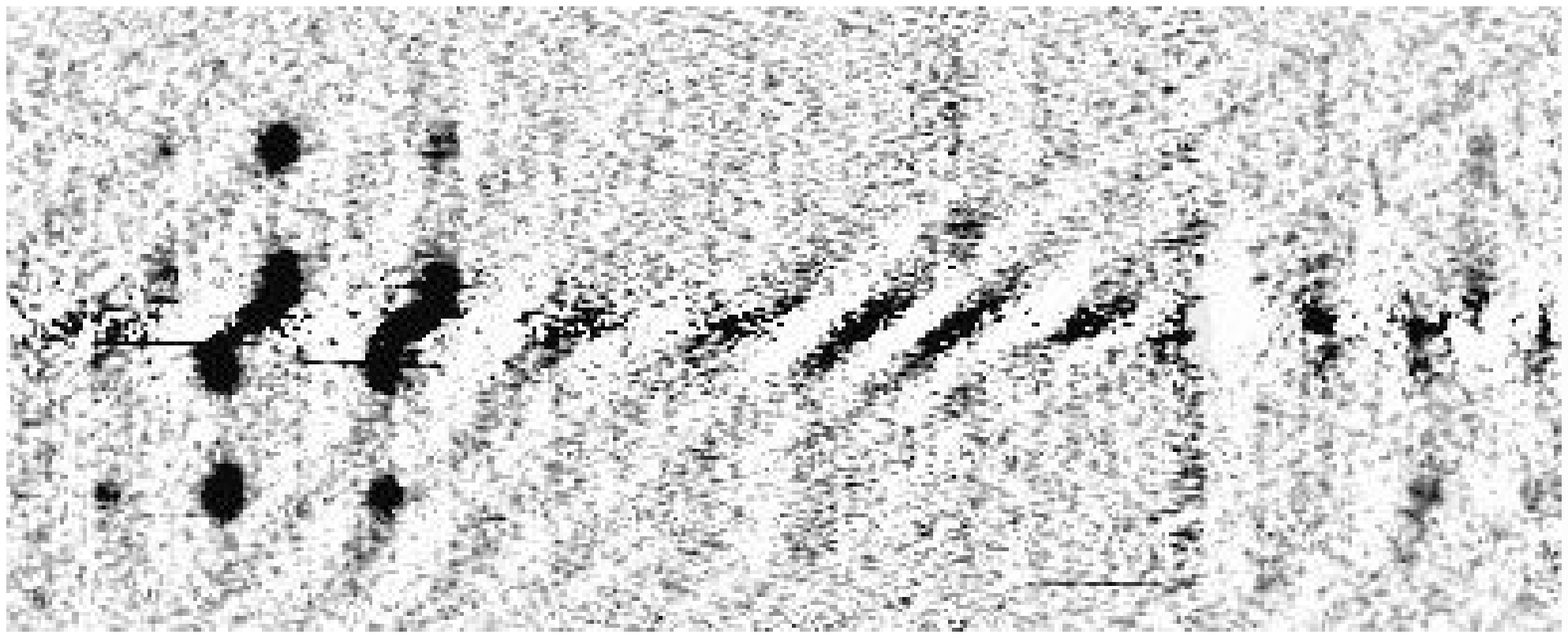}
   \includegraphics[width=0.45\textwidth]{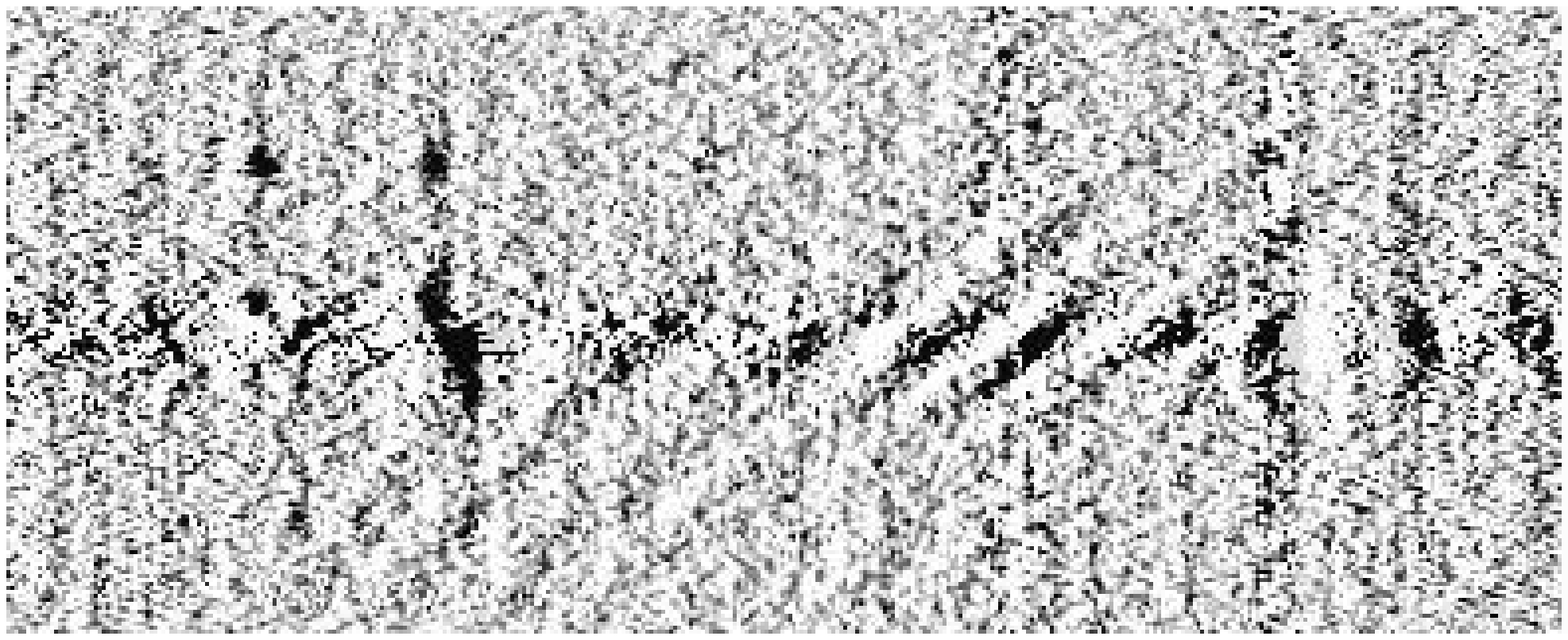}
   }
   \caption{The continuum-subtracted spectra near the \Ha\ emission line for NGC~6534 ({\it left}) and
   for MCG~11-22-015 ({\it right}).}
              \label{spectra}
    \end{figure*}
 
By inspecting the slit spectra of the galaxies, we have identified radius ranges where
the rings are betrayed by strong Balmer emissions; the spectra are shown in Fig.~\ref{spectra}.
We calculated the ratios of the strong emission lines to check the gas excitation. The excitation
can be checked with $\log (\mbox{[NII]}\lambda 6583 /\mbox{H}\alpha)$. In both cuts of the ring
in NGC~6534 and at the southern side of MCG~11-22-015 this ratio are
$-0.40 \pm 0.03$, $-0.37\pm 0.02$, and $-0.39\pm 0.20$, correspondingly.
In the ring of NGC~6534, for which we have also green spectrum, $\log (\mbox{[OIII]}\lambda 5007 /\mbox{H}\beta)$
is $-0.12\pm 0.15$ and $0.00\pm 0.09$. According to \citet{kewley06},
such emission-line ratios signify gas excitation by young stars in the southern parts of both rings and
possible composite excitation mechanism for the northern half of the NGC~6534 ring.
At the northern side of the MCG~11-22-015 where the emission lines are weak
$\log (\mbox{[NII]}\lambda 6583 /\mbox{H}\alpha) =-0.12\pm 0.12$
so here some shocks may be present.

When the gas is excited by young stars, one can estimate its oxygen abundance
by using empirical calibrations of the strong-line ratios. For the ring regions
where the gas is shown to be excited by young stars we have probed two
recognized calibrations -- that by \citet{pp04}, through N2 for NGC~6534
and MCG~11-22-015, and that by \citet{dopita16}, D16 with also the [SII]$\lambda$6717,6731 lines, for NGC~6534.
We have obtained $12+\log (\mbox{O/H}) =8.67\pm 0.02$ from N2 and
$12+\log (\mbox{O/H}) =8.65\pm 0.10$ from D16 for the southern tip
of the ring in NGC~6534; for the starforming site of the southern part of the ring in
MCG~11-22-015 the indicator N2 has given $12+\log (\mbox{O/H}) =8.68\pm 0.11$.
For the northern tip of the ring in NGC~6534 where the excitation may be composite the \citet{dopita16}
method gives $8.44\pm 0.06$. However, in any case the metallicity of the starforming gas
is comparable with the solar value or slightly lower.

\section{H$\alpha$ emission morphology and star formation rates in the rings}

  \begin{figure*}
%   \centering
%   \centerline{
%   \includegraphics[width=0.25\textwidth]{fig5a.eps} \qquad \qquad \qquad
%   \includegraphics[width=0.25\textwidth]{fig5b.eps}
%   }
%   \centerline{
%   \includegraphics[width=0.3\textwidth]{fig5c.eps} \qquad
%   \includegraphics[width=0.3\textwidth]{fig5d.eps}
%   }
%   \caption{The broad-band SDSS ({\it upper row}) and narrow-band H$\alpha$ ({\it bottom row})
%   images for NGC~6534 ({\it left}) and for MCG~11-22-015 ({\it right}).}
\resizebox{\hsize}{!}{\includegraphics[width=1.8cm]{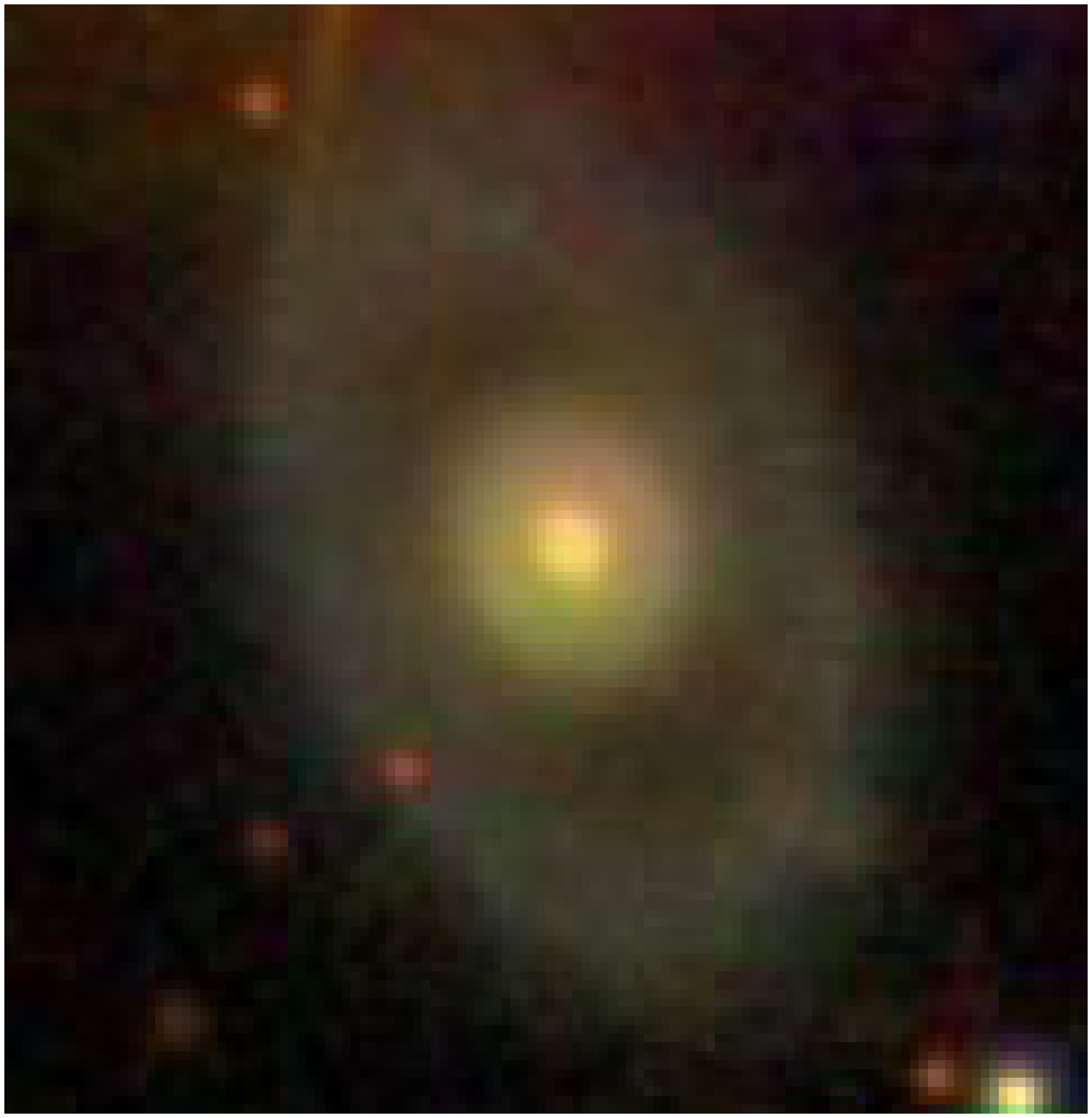}
\includegraphics[width=2.7cm]{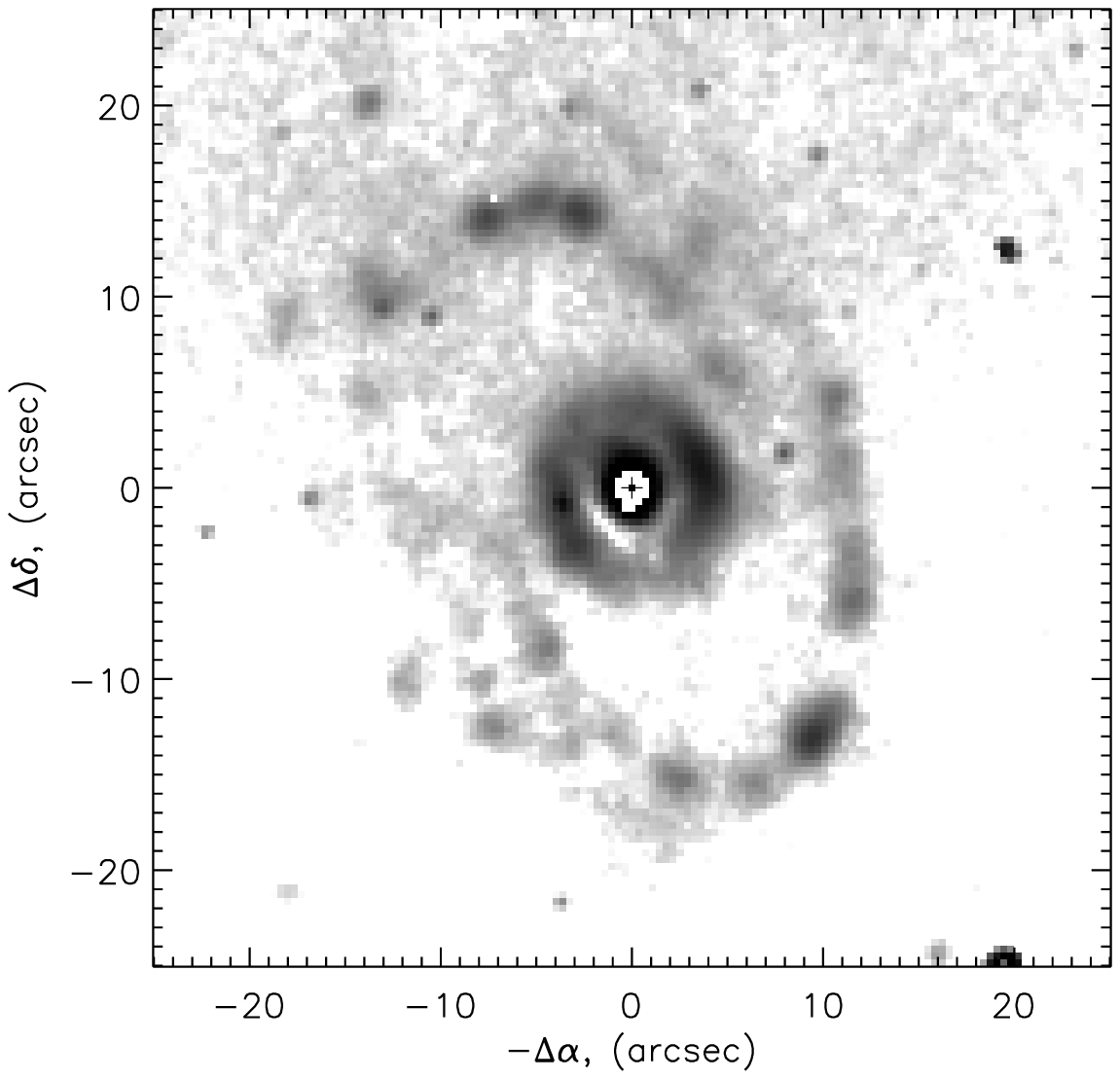}
\includegraphics[width=1.8cm]{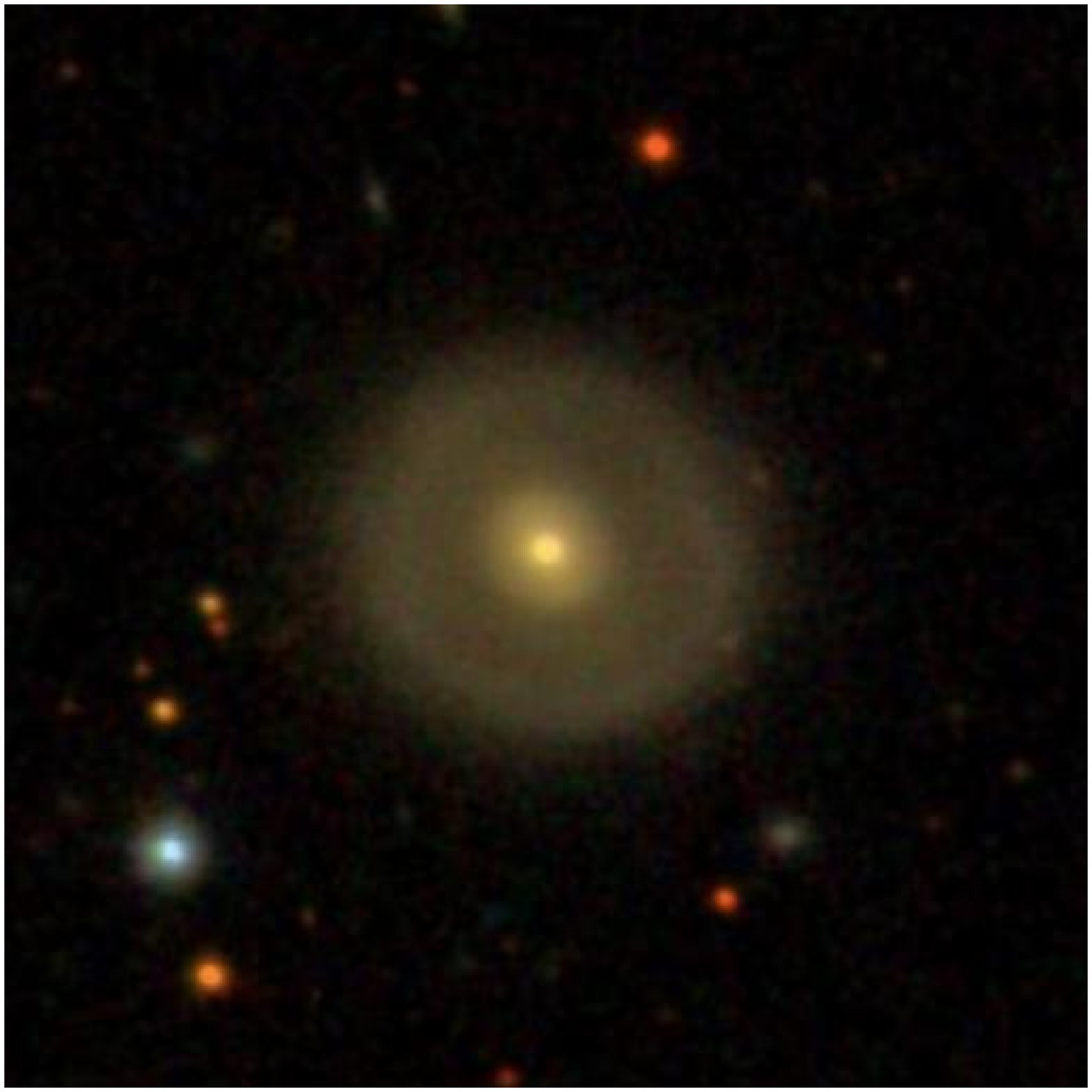}
\includegraphics[width=2.7cm]{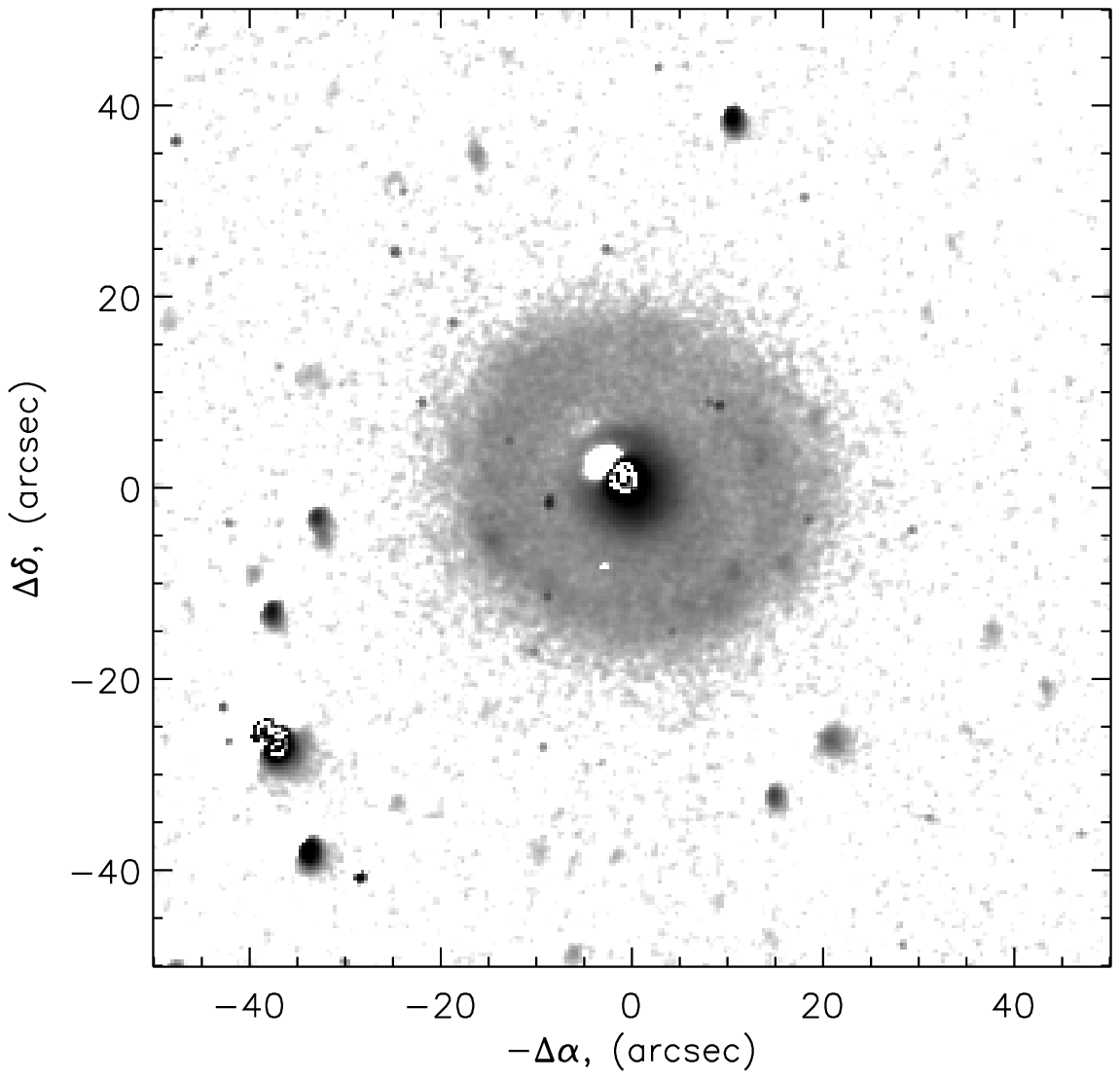}
}
\vspace{0.1cm}
\caption{The comparison of the broad-band SDSS images (left plot in every pair) and the narrow-band \Ha\
images (right plot in every pair).
NGC~6534 is to the left and MCG~11-22-015 is to the right.}
              \label{halpha}
    \end{figure*}

By undertaking narrow-band photometry of the galaxies through the filters
centered onto the redshifted H$\alpha$ line and onto the nearby continuum,
we have tried to subtract the latter images from the former ones to derive
net emission-line maps which are calibrated into energetic units
with the images of the spectrophotometric standard observed the same
night. While subtracting, the off-emission images were normalized to reduce
to zero the surrounding foreground stars implied to be pure-continuum sources.
The resulting net H$\alpha$ maps are shown in Fig.~\ref{halpha}.
The morphology of the emission-line image of NGC~6534 is somewhat unexpected:
while in the broad bands the galaxy has an evident ring, quite detached from the
main body of the galaxy, in the \Ha\ emission we see patchy spiral arms
starting from the ansae of the inner ring (Fig.~\ref{halpha}, left). The
\Ha\ morphology of MCG~11-22-015 is more traditional: it is a round, strongly
inhomogeneous ring, with its southern half more bright in the emission line;
it borders the inner lens, also filled by weak \Ha\ emission.

\begin{figure*}
   \centering
   \begin{tabular}{c c}
   \includegraphics[width=0.45\textwidth]{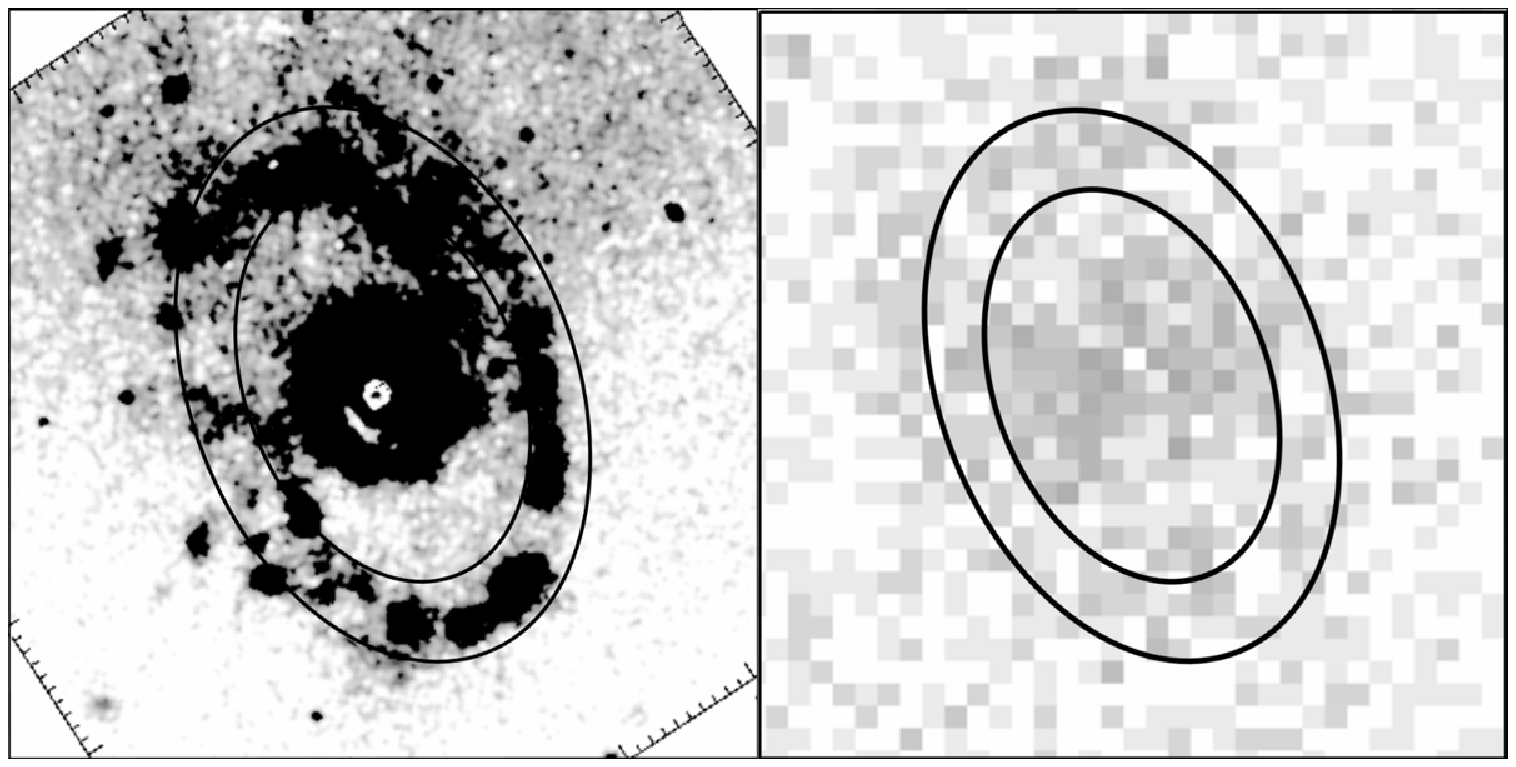} &
   \includegraphics[width=0.45\textwidth]{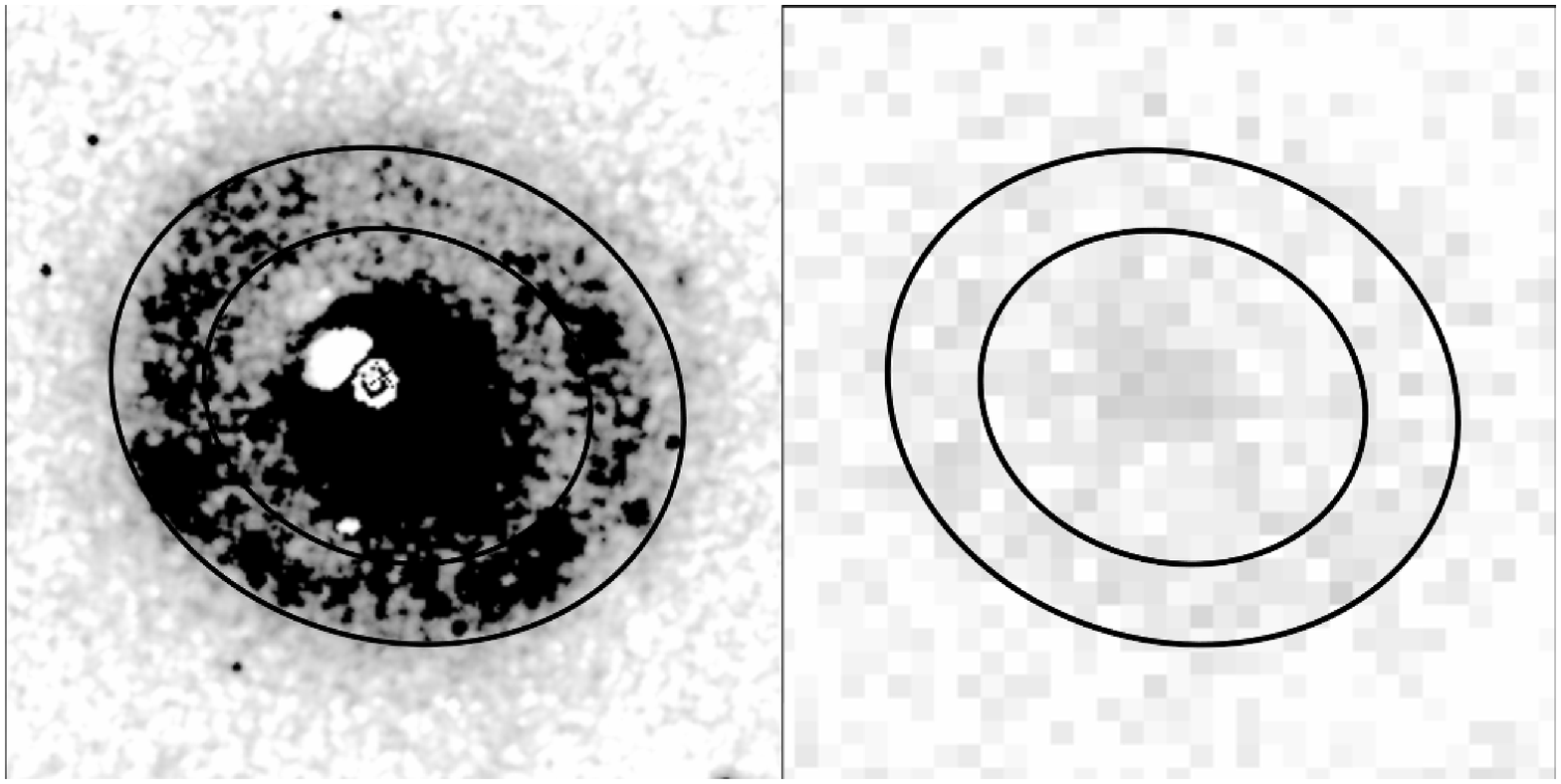} \\
   \end{tabular}
   \caption{The elliptical-annulus apertures used to integrate the fluxes for SFR
   derivations are overposed onto the H$\alpha$ and the NUV images (left and right plots in every pair).
   The size of the fields shown is 50\arcsec; NGC~6534 is to the left and MCG~11-22-015 is to the right.}
              \label{apertures}
    \end{figure*}

To calculate star formation rates (SFR) by using three independent indicators, namely,
ultraviolet, NUV and FUV fluxes from the GALEX images as well as our estimates
of the fluxes in the \Ha\ emission line, we have overposed the same elliptical
apertures onto all 3 images of every galaxy. To determine the borders of the apertures, we have
used the H$\alpha$ images because they are the ones with the most contrast. The apertures
are shown in Fig.~\ref{apertures}. We have measured the fluxes inside the elliptical annuli,
just between the inner and outer ellipses of Fig.~\ref{apertures} for the three above-mentioned bands
for NGC~6534 and only for the deep NUV/MIS image
and the H$\alpha$ image of MCG~11-22-015 -- the latter galaxy is weak in the UV, and
its shallow FUV/AIS image does not contain any signal in the ring above $3\sigma$
of the sky level. Then we have applied the calibrations of the SFR from \citet{ken_evans}
and have corrected the UV-based estimates of SFR for the intrinsic dust
as it is described in \citet{kostuk18}, by using the WISE imaging in 22$\mu$m.

We found that over the NGC~6534 ring the star formation proceeded with a roughly constant rate
during the last 100--200~Myr: the SFRs from the NUV, FUV (GALEX/AIS), and H$\alpha$
fluxes are 0.13, 0.12, and 0.18~\SFR, respectively, with an individual accuracy of the estimates
of 0.03~\SFR.

In MCG~11-22-015 the situation is more complex.
The NUV flux measured in the deep MIS-image gives SFR$=0.07\pm 0.01$~\SFR, while the \Ha\ flux
gives SFR$=0.26\pm 0.01$~\SFR\ with the standard
calibration. In this situation we remember the caveat from \citet{calzetti13}
that the standard UV-SFR calibration is valid only for star formation proceeding over
long timescales, $\sim 100$~Myr. If the SF timescale is shorter, we must increase
the SFR estimate; for example, for SF occuring during only the last 2~Myr we must
multiply the standard calibration by a factor of 3.45 \citep{calzetti13}.
By multiplying our UV-based SFR of 0.07~\SFR\ by 3.45, we are obtaining
0.24~\SFR. The latter estimate leads to consistency with the SFR from the \Ha.
So for MCG~11-22-015 we may suggest that its star formation in the ring
is quite recent -- no older than a few Myr.

\section{Conclusions}

   We have studied the outer starforming rings in two lenticular
galaxies, NGC~6534 and MCG~11-22-015: spectral data as well as narrow-band
and UV images are analyzed.  The ionized gas in the rings is mostly excited by young stars
and demonstrates nearly solar metallicity rejecting a possibility of its origin
from cosmological filaments. Despite the classification
of NGC~6534 as an isolated galaxy in the NED, we have detected a very close companion,
PGC~2666218, at some 40~kpc from it in projection, with the similar radial velocity. MCG~11-22-015 has already
been known as a member of an X-ray bright group \citep{group_rosat}. So there is no problem with
the outer gas sources feeding star formation -- it may be tidal harassment of the neighbour by NGC~6534 and
a minor merger for MCG~11-22-015. Star formation histories in the rings are also
different: the SFR has been constant at least for the last 200~Myr in NGC~6534, in
accordance with the possible resonance nature of its ring, and it started recently
in MCG~11-22-015 betraying a fresh gas-rich satellite cannibalism. The quasi-solar, or only
slightly subsolar metallicity of the gas in the outer starforming rings of S0 galaxies was found by us
previously more than once, including the cases of obvious minor mergers -- see, for example,
NGC~4513 with its counterrotating gas \citep{we_uvrings}. Perhaps, we need some specific effective
chemical evolution mechanisms to explain this phenomenon.

\begin{acknowledgements}
      The study of the galactic rings was supported by the Russian Foundation for Basic
      Researches, grant no. 18-02-00094a. The work is based on the data obtained at the Russian
      6m telescope of the Special Astrophysical Observatory and on the public data
      of the SDSS (http://www.sdss3.org) and GALEX (http://galex.stsci.edu/GR6/) surveys.
      We are grateful to D. Oparin, A. Moiseev, and
      O. Egorov having supported our observations at the Russian 6m telescope (operating under
      the financial support of the Ministry of Science and Education of the Russian Federation,
      agreement No14.619.21.0004, the project RFMEFI61914X0004).
      This research has made use of the NASA/IPAC Extragalactic Database (NED, http://ned.ipac.caltech.edu)
      which is operated by the Jet Propulsion Laboratory, California Institute of Technology,
      under contract with the National Aeronautics and Space Administration, and of
      the Lyon Extragalactic Database (HyperLEDA, http://leda.univ-lyon1.fr).
\end{acknowledgements}


\end{document}